# Mott Relation for Anomalous Hall and Nernst effects in $Ga_{1-x}Mn_xAs$ Ferromagnetic Semiconductors

Yong Pu[1], Daichi Chiba[2], Fumihiro Matsukura[2], Hideo Ohno[2] and Jing Shi[1]

[1] Department of Physics and Astronomy, University of California, Riverside, CA 92521

[2] Semiconductor Spintronics Project, ERATO-JST, Sendai, Japan, and Laboratory of Nanoelectronics and Spintronics, Research Institute of Electrical Communication, Tohoku University, Sendai, Japan

The Mott relation between the electrical and thermoelectric transport coefficients normally holds for phenomena involving scattering. However, the anomalous Hall effect (AHE) in ferromagnets may arise from intrinsic spin-orbit interaction. In this work, we have simultaneously measured AHE and the anomalous Nernst effect (ANE) in $Ga_{1-x}Mn_xAs$ ferromagnetic semiconductor films, and observed an exceptionally large ANE at zero magnetic field. We further show that AHE and ANE share a common origin and demonstrate the validity of the Mott relation for the anomalous transport phenomena.

Topological or dissipation-less spin current, or the intrinsic spin Hall effect, in *p*-type GaAs semiconductor was recently proposed for room-temperature spin sources for spintronics[1] owing to the spin-orbit interaction. Spin accumulation, a consequence of the spin Hall effect, has been experimentally observed in GaAs.[2,3] In diluted magnetic semiconductors (DMS) such as $Ga_{1-x}Mn_xAs$, the spin-polarized holes also experience the same spin-orbit interaction in addition to the random scattering potentials, resulting in a related effect, i.e. a net transverse charge current or the Hall current. As in many other ferromagnets, this well-known anomalous Hall effect (AHE) has been routinely employed to characterize the magnetic properties of DMS films,[4] for its magnitude is directly proportional to the magnetization. Several models based on the intrinsic (e.g. inter-band effect[5] and the Berry phase[6]) and extrinsic (skew scattering[7] and side-jump[8]) mechanisms have been put forward to account for AHE. The intrinsic and side-jump mechanisms give rise to an anomalous Hall current $\vec{J}_H$, a scattering-rate ($\frac{1}{\tau}$)-independent anomalous Hall conductivity $\sigma_{xy}^{AH}$. This so-called dissipation-less AHE can be paraphrased by a power-law relation between the anomalous Hall resistivity $\rho_{xy}^{AH}$ and the longitudinal resistivity $\rho_{xx}$ with exponent *n*=2, i.e. $\rho_{xy}^{AH} \sim \rho_{xx}^2$; because $\sigma_{xy}^{AH} = \frac{-\rho_{xy}}{\rho_{xx}^2 + \rho_{xy}^2} \approx -\frac{\rho_{xy}}{\rho_{xx}^2}$, $\sigma_{xy}^{AH}$ is independent of $\frac{1}{\tau}$. In contrast, the skew scattering mechanism predicts *n* =1, *i.e.* a $\frac{1}{\tau}$-dependent $\sigma_{xy}^{AH}$. A possible crossover between these two regimes was recently proposed for two-dimensional systems.[9]

In $Ga_{1-x}Mn_xAs$ it was theoretically shown that the Berry phase acquired by quasi-particles moving on the spin-split Fermi surfaces is responsible for the observed magnitude of AHE.[6] The same theory also satisfactorily explained the observed anisotropic magnetoresistance in $Ga_{1-x}Mn_xAs$.[10] The evidence appears to suggest a predominate role of the intrinsic spin-orbit interaction. An experimental attempt to test the power-law scaling in Ga$_{1-x}$Mn$_x$As was made by Edmonds et al.,[11] but the range in $\rho_{xx}$ was very limited. In a more recent study by Chun et al.,[12] $\rho_{xx}$ spans over a greater range by varying Mn concentration similar to an earlier study in CuCr$_2$Sr$_{4-x}$Br$_x$[13], and $n = 2$ was found in the metallic regime. Because the side-jump was believed to play a negligibly small role in DMS, they concluded that the intrinsic mechanism prevails in the metallic regime.[12]

The anomalous Nernst effect (ANE) is the thermoelectric counterpart of AHE. In general, the Seebeck coefficient $S$ is related to the energy (E) derivative of the electrical conductivity $\sigma$ at the Fermi level, through the well-known Mott's relation, *i.e.* $S = \frac{\pi^2 k_B^2 T}{3e\sigma}\left(\frac{\partial \sigma}{\partial E}\right)_{E_F}$. Along with the Onsager reciprocal relations, the Mott relation is another general relation linking different transport coefficients.[14] In the presence of a magnetic field, the Mott relation also holds for the off-diagonal elements of the transport coefficient tensors.[15] Lee *et al.* applied the Mott relation to spinel[16] in the case of dissipation-less transport ($n = 2$). It was not entirely clear whether the Mott relation is applicable to the dissipation-less AHE especially the intrinsic AHE, until a theoretical proof was given recently by Xiao et al.[17] Experimentally, ANE study in DMS has not yet been reported and, moreover, the Mott relation has not been firmly established in any ferromagnet. It is the objective of this work to show the scattering independent nature of both AHE and ANE in Ga$_{1-x}$Mn$_x$As, and furthermore to validate the Mott relation for the anomalous Hall and Nernst effects.

In most experimental studies,[11,13,18] magnetic films have in-plane anisotropy, thus the Hall voltage is zero unless a perpendicular magnetic field ***B*** is applied to rotate the magnetization ***M*** out of the plane. Since AHE is proportional to the out-of-plane magnetization, the rotation of ***M*** produces a large field-dependent AHE signal at relatively low fields. To accurately test the power-law, a large ***B***-field in excess of several teslas must be applied to fully saturate **M** along **B**. However, at sufficiently high **B**-fields when **M** is fully saturated, $\rho_{xy}$ does not saturate due to the ordinary Hall effect (linear in **B**) and the magnetoresistance in $\rho_{xx}$ that may also be a strong function of **B**. In order to separate out the two effects, one needs to know the precise relationship between $\rho_{xy}$ and $\rho_{xx}$ as **B** is varied. Unfortunately, $\rho_{xy}$ and $\rho_{xx}$ do not follow the same power-law predicted for zero magnetic field. As evidenced in Fig. 1a for GaMnAs (x=0.05), $\rho_{xx}$ changes by ~50%, whereas $\rho_{xy}$ changes by only ~20% over the same field range. This complication makes the accurate extrapolation of $\rho_{xy}$ to zero field unreliable or futile. To overcome this difficulty, we have thus prepared $Ga_{1-x}Mn_xAs$ films with perpendicular anisotropy. From the squared hysteresis loops, we can readily determine the *spontaneous* AHE and ANE coefficients; therefore, the power-law can be tested using transport coefficients all measured precisely at **B** =0.

Four 50 nm-thick $Ga_{1-x}Mn_xAs$ (*x* =0.04-0.07) DMS films were grown by molecular beam epitaxy. To engineer the perpendicular anisotropy needed for this study, a 500 nm-thick (Ga, In)As buffer layer was used to produce tensile strain. The Curie temperature $T_c$ of the as-grown samples ranges from 70 to 110 K. Three samples were annealed at 250 $^\circ$C in air for 60 minutes. Upon annealing, $T_c$ is dramatically increased, accompanied by a decrease in resistivity and a slight increase in the coercive field. For comparison, we also measured one sample (x=0.05) in the pre-annealed state. The films were patterned into the Hall bars along [110]. Simultaneous electrical and thermoelectric measurements were carried out in a continuous-flow liquid helium cryostat with an electromagnet (up to 0.2 T) described previously[19], and the high-field resistivity measurements were carried out in a physical property measurement system (PPMS). At each temperature and magnetic field, $\nabla T$ is set up along the Hall-bar length by turning on a heater at one end of the sample, and a linear temperature distribution is assumed along the sample. Both longitudinal and transverse open-circuit thermal emf voltages $V_x$ and $V_y$ are recorded simultaneously. From $V_x$, $V_y$, and $\nabla T$, the Seebeck coefficients $S_{xx}$ and $S_{yx}$ are obtained using $S_{xx} = E_x/(\nabla T)_x$ and $S_{yx} = E_y/(\nabla T)_x$. All four coefficients, $\rho_{xx}$, $\rho_{xy}$, $S_{xx}$ and $S_{xy}$ (= - $S_{yx}$) are measured as $B$ is swept at each $T$. In addition, the magnetic hysteresis loops and $T$-dependence of $M$ are measured using a SQUID magnetometer.

We first focus on the resistivity. As shown in Fig. 1b, $\rho_{xy}$ exhibits a squared hysteresis as expected for films with robust perpendicular anisotropy. Since $\rho_{xy} = R_H B + \rho_{xy}^{AH}(M)$, where the first- and second-terms are the ordinary and anomalous Hall resistivities respectively, in the following quantitative analysis we have $\rho_{xy} = \rho_{xy}^{AH}$ since we always use the zero-field values.

In similar analyses adopted by other researchers, whether $T$, $B$, or impurity concentration $x$ is chosen as the controlling parameter, one needs to measure three quantities: $\rho_{xx}$, $\rho_{xy}$ and $M_z$, in two independent experiments. As seen in Fig. 1c, the $M$ vs. $B$ loop measured by SQUID differs quite significantly from the $\rho_{xy}$ vs. $B$ loop. Whereas the $\rho_{xy}$ vs. $B$ loops in all samples are always squared at low $T$, the $M$ vs. $B$ loops are rarely so. This key difference lies in what $\rho_{xy}$ and $M$ actually measure. $\rho_{xy}$ is picked up wherever there is a current in the sample. In DMS, spins in carrier-rich regions where the conductivity is relatively higher contribute to both $\rho_{xy}$ and $M$. On the contrary, spins in isolated regions do not contribute at all to $\rho_{xy}$, but do contribute to $M$. Additionally, $\rho_{xy}$ and $M$ measurements are usually taken on separate samples of drastically different sizes. The former is from a small Hall cross but the latter is from a sample about 2-3 orders larger. Not surprisingly, using $M_Z(T)$ from SQUID measurements may cause errors in the power-law analysis. To circumvent this problem, we take advantage of the second pair of transport coefficients, i.e. $S_{xx}$ and $S_{xy}$ from the thermoelectric measurements in which the signals come from the exactly same conducting regions over the same Hall-cross area.

As shown in the left column of Fig. 2, ANE loops (red) are plotted together with the simultaneously measured AHE loops (blue) and the two sets of loops match exceedingly well (up to a scaling factor). The well-matched AHE and ANE loops are the evidence that both effects scale with $M$ in the same fashion. Displayed in the right column are both AHE and ANE loops for $x=0.04$ annealed sample (labeled as 0.04*) measured at different temperatures. The figure shows a striking contrast between these two effects: ANE changes the sign at some intermediate $T$, whereas AHE remains positive at all temperatures. The sign change in $S_{yx}$ occurs in all three annealed samples. As $T$ approaches $T_c$, the loops narrow atop a smooth background, which resembles the shape of $M$ due to diminished magnetic anisotropy. However, the nearly perfect match between the AHE and ANE loops suggests that AHE and ANE follow the identical $M$-dependence; therefore, share a common physical origin.

The robust perpendicular anisotropy allows us to take the zero-field value of $S_{yx}$ for further analysis. Both zero-field $S_{xx}$ and $S_{yx}$ are shown in Fig. 3 as functions of $T$. $S_{xx}$ is always positive as expected for $p$-type semiconductors. Note that in all three annealed samples, a high peak emerges at low temperatures. Similar annealing effect on $S_{xx}$ was also observed in other in-plane anisotropy GaMnAs samples. $S_{yx}$ is zero above $T_c$, consistent with the fact that ANE is proportional to the spontaneous magnetization. Unlike AHE that remains finite as T→ 0, ANE goes to zero as the entropy should vanish at $T=0$. Although the physical origin of the $S_{xx}$ peak in annealed samples remains a subject of further investigation, we attribute it to the enhanced phonon-drag resulting from the improved phonon mean-free-path in annealed samples. Regardless of the underlying mechanism, in the following, we point out that the occurrence of the peak in $S_{xx}$ at low T can be correlated with the sign change in $S_{yx}$.

The Seebeck coefficient is related to other transport coefficients by

$$S_{yx} = \frac{1}{\sigma_{xx}}(\alpha_{yx} - \sigma_{yx} S_{xx}) \quad (1)$$

where $\alpha_{yx}$ is the Nernst conductivity defined by $\vec{J} = \sigma\vec{E} + \alpha(-\nabla T)$, where $\vec{J}$ is the electric current density and $\vec{E}$ is the electric field. Hence, the Nernst effect can be quantitatively represented by either $S_{yx}$ or $\alpha_{yx}$. From $S_{xx}$, $S_{yx}$, $\sigma_{xx}$ and $\sigma_{yx}$, we can determine $\alpha_{yx}$ according to Eq. 1, as shown in Fig. 4. On the other hand, if the Mott relation holds, $\alpha_{yx}$ can be calculated from $\sigma_{yx}$ via $\alpha_{yx} = \frac{\pi^2 k_B^2 T}{3e}\left(\frac{\partial \sigma_{yx}}{\partial E}\right)_{E_F}$, where E is the energy. If we straightforwardly adopt the power-law, $\rho_{xy}^{AH} = \lambda M_z \rho_{xx}^n$ (assuming an arbitrary exponent $n$), and substitute it into the Mott relation, then we can easily see that both sides of Eq. 1 contain a common factor, $M_z$! Since the exactly *same* magnetization from the identical part (i.e. the Hall cross) of the sample contributes to both AHE and ANE as discussed earlier, $M_z$ disappears from the following two equivalent equations (Eqs. 2 and 3). As a result, the critical test of the Mott relation does not need to involve any magnetization, but only the four transport coefficients,

$$S_{yx} = \frac{\rho_{xy}}{\rho_{xx}}\left(T\frac{\pi^2 k_B^2}{3e}\frac{\lambda'}{\lambda} - (n-1)S_{xx}\right) \quad (2) \text{ and } \alpha_{yx} = \frac{\rho_{xy}}{\rho_{xx}^2}\left(T\frac{\pi^2 k_B^2}{3e}\frac{\lambda'}{\lambda} - (n-2)S_{xx}\right) \quad (3).$$

The pre-factor $\lambda$ in the power-law has to depend on the Fermi energy of the hole gas; otherwise its energy derivative $\lambda'$ would vanish, which leads to zero $\alpha_{yx}$. If the power-law is obeyed as $T$ is varied, it means that neither $\lambda$ nor $\lambda'$ depends on $T$. Recall that $\rho_{xx}$, $\rho_{xy}$ or $S_{xx}$ does not change sign over the entire $T$-range. From Eq.2 and Fig. 3a, we immediately conclude that the sign change in $S_{yx}$ would not be possible if $n =1$. Consequently, the sign change alone allows us to exclude the possibility of the skew-scattering mechanism for AHE. As $T$ is lowered from $T_c$, the importance of the first-term steadily diminishes; however, in the $T$-window where $S_{xx}$ shows a peak, since $S_{xx}$ is always positive in our *p*-type samples, the second-term can take over to cause a sign change only if n>1. It is also enlightening to examine Eq. 3. We know in Fig. 4 that the measured $\alpha_{yx}$ remains positive over the whole $T$-range; therefore, from Eq. 3, we find that $n$ cannot be greater than two at least for the annealed samples.

To find the exponent, we treat $\lambda'/\lambda$ and $n$ as two fitting parameters. By fitting Eq. 2 to $S_{yx}$ (Fig. 3a) or Eq. 3 to $\alpha_{yx}$ (Fig. 4) for all samples, we can search for the best-fit values for $n$ and $\lambda'/\lambda$. Both fits should yield the same set of values. In fact, the solid curves in Fig. 3a and Fig. 4 are the best fits with essentially the same fitting parameters. Obviously, the fits not only capture the sign change and curvature changes in $S_{yx}$, but also work very well for both $S_{yx}$ and $\alpha_{yx}$ over the entire $T$-range. This unmistakably demonstrates the validity of the Mott relation for AHE and ANE. Moreover, as seen in Fig. 4, the best-fit exponent is very close to two for all samples. It proves that AHE is scattering-independent in all GaMnAs samples. It should be emphasized that the determination of $n$ only requires four transport coefficients from the exactly same area of the sample, which removes any possible uncertainty introduced in the $M$-measurements. This gives us a strong sense of confidence in the determination of the exponent, and therefore the scattering-independent nature of AHE. In DMS, intrinsic mechanism was shown to be a dominant mechanism[6, 10, 12], therefore, n=2 strongly favors the intrinsic mechanism in DMS. For the intrinsic AHE ($n$ =2), the second-term in Eq. 3 vanishes, indicating that $\alpha_{yx}$ does not at all depend on $S_{xx}$, or has nothing to do with scattering. In other words, $\alpha_{yx}$ depends only on the electronic band structure and the magnetization of the samples, implying an intrinsic Nernst current, $\vec{J}_N = \alpha(-\nabla T)$. To reinforce this point, in Fig. 4, we also plot the fitting curves (dashed lines) with $n$ fixed at one. Obviously the single-variable fitting is not possible in annealed samples. Clear peak features in $S_{xx}$ can be mirrored in those $\alpha_{yx}$ curves (dashed), illustrating that an extrinsic (*i.e.*

n=1) Nernst current would strongly depend on $S_{xx}$.

In summary, we have demonstrated the intrinsic origin of AHE and ANE in Mn-doped GaAs ferromagnetic semiconductors. From four transport coefficients measured at zero magnetic field, we have verified the Mott relation for the off-diagonal transport coefficients for the intrinsic mechanism.

J.S. and Y.P. sincerely thank Q. Niu and D. Xiao for many stimulating discussions and acknowledge the support of DOE and CNID. A part of the work at Tohoku University was supported by the GCOE program and by Gran-in-Aids from MEXT/JSPS.

Figure captions:

**Fig. 1.** (Color online) (a) $\rho_{xy}$ and $\rho_{xx}$ vs. B for 5% sample at T= 2 K. The lower panel shows the log-log plot of the same $\rho_{xy}$ and $\rho_{xx}$ data (symbols) and three dashed lines correspond to the power-law with different exponents in $\rho_{xy} \propto \rho_{xx}^n$: n=2, 1 and 0.5. (b) Zero-field $\rho_{xy}$ (upper) and $\rho_{xx}$ (lower) for all four samples. The annealed samples are denoted by "*". The inset in the upper panel is an AHE loop of 7% annealed sample (x=0.07*) at 10 K. (c) M and $\rho_{xy}$ loops of x=0.07* sample at 8 K.

**Fig. 2.** (Color online) AHE and ANE loops at T=10K for different samples (left) and at different temperatures for 4% annealed sample (right). In the left panel, ANE data of 0.04*, 0.05* and 0.07* samples were multiplied by -1.

**Fig. 3.** (Color online) Temperature dependence of $S_{yx}$ (a) and $S_{xx}$ (b) for all samples measured at zero magnetic field. In (a), the solid lines are the best fits using Eq. 2.

**Fig. 4.** (Color online) Zero-field Nernst conductivity $\alpha_{yx}$ for all samples. The solid red lines are the best fits using Eq. 3 and dashed curves are the best fits with *n=1*.

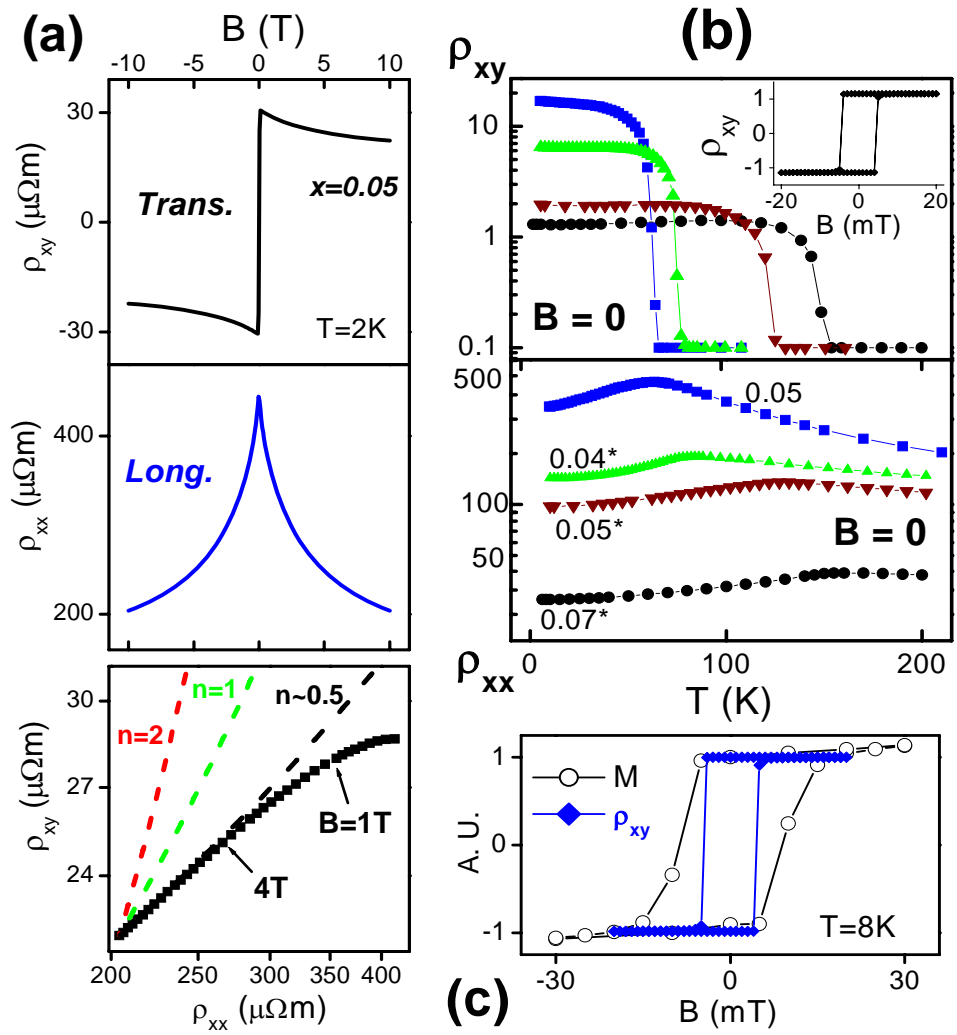

Figure 1

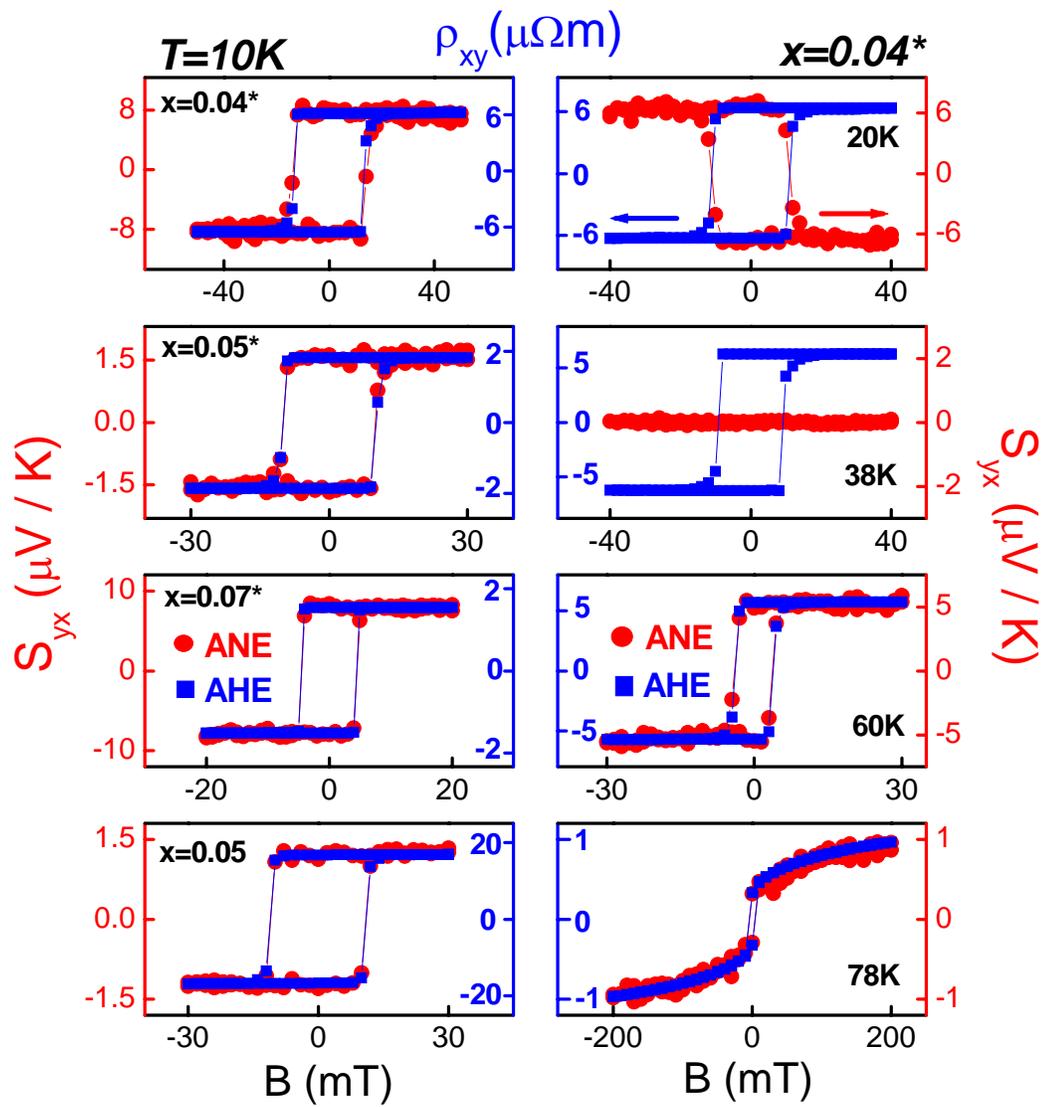

Figure 2

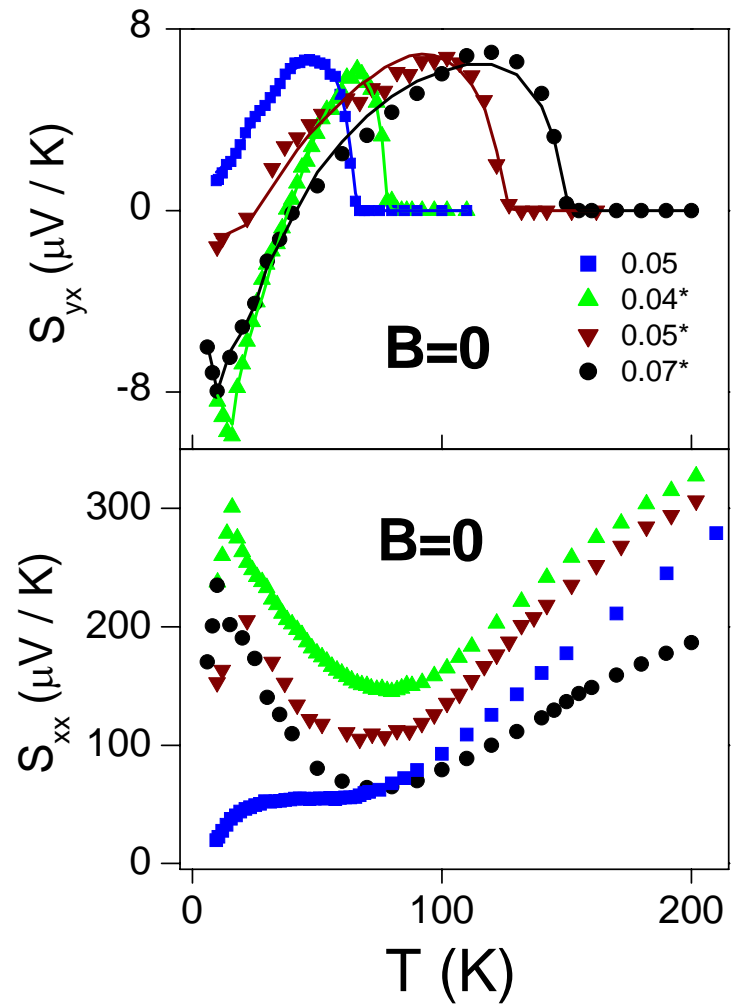

Figure 3

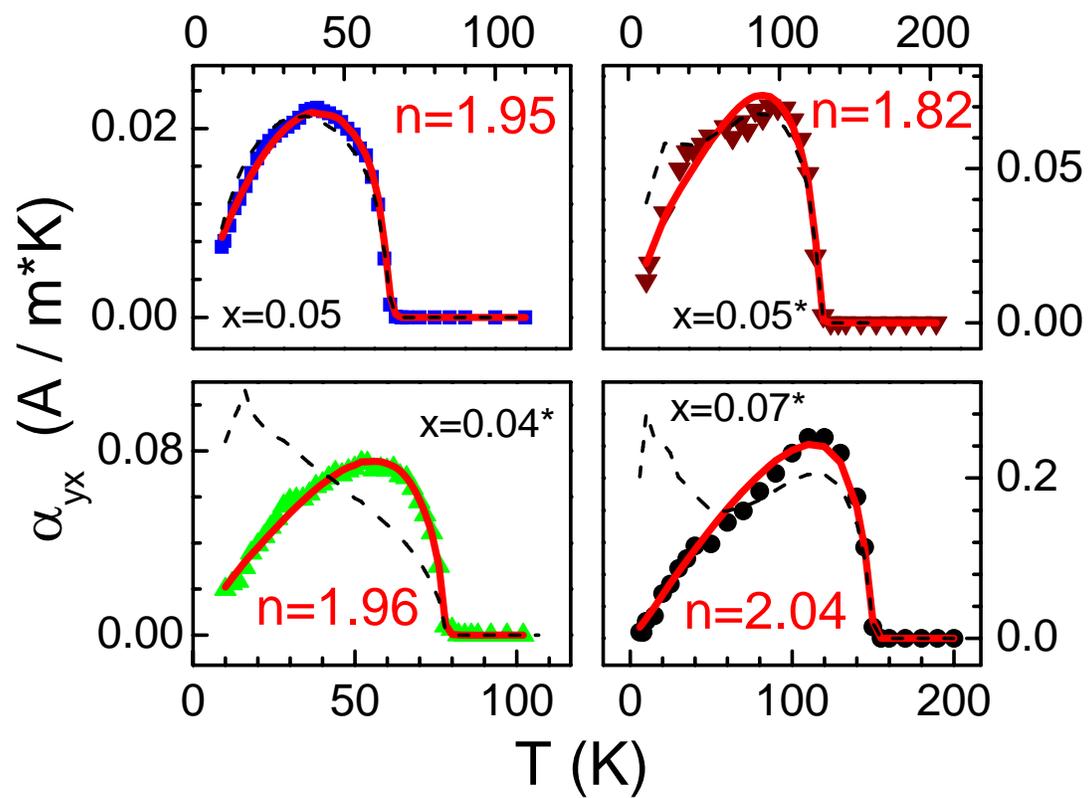

Figure 4

References:

[1] S. Murakami, N. Nagaosa, S.-C. Zhang, *Science* **301**, 1348 (2003).

[2] Y.K. Kato et al., *Science* **306**, 1910 (2004).

[3] J. Wunderlich et al., *Phys. Rev. Lett.* **94**, 047204 (2005).

[4] H. Ohno, *Science* **281**, 951 (1998).

[5] R. Karplus, J.M. Luttinger, *Phys. Rev.* **95**, 1154 (1954).

[6] T. Jungwirth, Q. Niu, A.H. MacDonald, *Phys. Rev. Lett.* **88**, 207208 (2002).

[7] J. Smit, Physica (*Amsterdam*) **21**, 877 (1955).

[8] L. Berger, *Phys. Rev. B* **2**, 4559 (1970).

[9] S. Onoda et al., *Phys. Rev. Lett.* **97**, 126602 (2006).

[10] T. Jungwirth et al., *Appl. Phys. Lett.* **83**, 320 (2003).

[11] K.W. Edmonds et al., *J. Appl. Phys*. **93**, 6787 (2003).

[12] S.H. Chun et al. Phys. Rev. Lett. **98**, 026601 (2007).

[13] W.-L. Lee et al., *Science* **303**, 1647 (2004).

[14] N.F. Mott, H. Jones, *The Theory of the Properties of Metals and Alloys* (Dover Publications, Inc. New York, 1958).

[15] Y. Wang et al., *Phys. Rev B* **64** 224519 (2001).

[16] W.-L. Lee et al., *Phys. Rev. Lett.* **93**, 226601 (2004).

[17] D. Xiao, Y. Yao, Z. Fang, Q. Niu, *Phys. Rev. Lett.* **97**, 026603 (2006).

[18] C. Zeng et al., *Phys. Rev. Lett.* **96**, 037204 (2006).

[19] Y. Pu et al., *Phys. Rev. Lett*. **97**, 036601 (2006).